\documentclass[10pt,
               aps,
               prl,
               twocolumn,
               showpacs, 
                nolongbibliography,
               groupedaddress]{revtex4-2}
\usepackage[utf8]{inputenc}
\usepackage{graphicx}
\graphicspath{{images/}}
\usepackage[caption=false]{subfig}
\usepackage{amsmath, amsfonts, amssymb}
\usepackage{hyperref}
\usepackage{tikz}
\makeatletter
\newcommand*{\rom}[1]{\expandafter\romannumeral #1}
\makeatother

\begin{document}

\title{Random self-propulsion to rotational motion of a microswimmer with inertial memory}

\author{F Adersh\textsuperscript{$\dag$}}

\affiliation{Department of Physics, University of Kerala, Kariavattom, Thiruvananthapuram-$695581$, India}

\author{M Muhsin\textsuperscript{$\dag$}}

\affiliation{Department of Physics, University of Kerala, Kariavattom, Thiruvananthapuram-$695581$, India}

\author{M Sahoo}
\email{jolly.iopb@gmail.com}
\affiliation{Department of Physics, University of Kerala, Kariavattom, Thiruvananthapuram-$695581$, India}

\date{\today}

\begin{abstract}
We study the motion of an inertial microswimmer in a non-Newtonian environment with a finite memory and present the theoretical realization of an unexpected transition from its random self-propulsion to rotational (circular or elliptical) motion. Further, the rotational motion of the swimmer is followed by spontaneous local direction reversals yet with a steady state angular diffusion. Moreover, the advent of this behaviour is observed in the oscillatory regime of the inertia-memory parameter space of the dynamics. We quantify this unconventional rotational motion of microswimmer by measuring the time evolution of direction of its instantaneous velocity or orientation. By solving the generalized Langevin model of non-Markovian dynamics of an inertial active Ornstein-Uhlenbeck particle, we show that the emergence of the rotational (circular or elliptical) trajectory is due to the presence of inertial memory in the environment or medium.
\end{abstract}

\maketitle
Microswimmers are typical nonequilibrium systems which are driven away from equilibrium due to their ability to utilize energy obtained from the surroundings and convert it into self-propelled motion~\cite{elgeti2015physics, menzel2015tuned}. Examples of microswimmers include flagellated bacteria, eukaryotic cells, sperms, ciliates, synthetic microswimmers and so on~\cite{ali2012bacterial, corbyn2021stochastic, vig2012swmming, kruger2016curling, howse2007self}. Though the propulsion of microswimmers has been widely explored, most of the literature disregards the influence of memory of the medium in which they propel. In a practical scenario, microswimmers are mostly exposed to non-Newtonian fluids like polymer solution~\cite{zottl2019enhanced, qi2020enhanced, liu2021viscoelastic}, crystalline medium~ \cite{zhou2014living, ferreiro2018long} or biological environments such as mucus~\cite{mirbagheri2016helicobacter}, tissues etc., where memory plays a significant role in their dynamics~\cite{plan2020active, teran2010viscoelastic, shen2011undulatory}.
    Due to the presence of memory in the environment, an enhancement in the attraction and orientational ordering of the swimmers~\cite{li2016collective},
    an increase in rotational diffusion of Janus particles with self-propulsion velocity~\cite{solano2016dynamics}, an enhancement of swimming speed of  \textit{E. Coli} bacteria with 
    reduced tumbling~\cite{binagia2021self}, etc are observed.
    
Moreover, rotation plays a crucial role in the dynamics of microswimmers~\cite{lauga2005swimming,shenoy2007kinematic, takagi2013dispersion}. It enables the microswimmers to avoid obstacles, locate food and other resources, and interact with their surroundings. In case of chiral microswimmers with shape asymmetry, circular trajectory is a typical feature of their motion~\cite{kummel2013circular, lowen2016chirality}. The Chiral active colloid which is a kind of microswimmer undergoes a velocity dependent viscous torque that yields circular trajectories~\cite{takagi2013dispersion, kummel2013circular}. However, chirality is not the only key factor for the occurrence of rotational motion of microswimmers. Achiral microswimmers can also make a transition from random self-propulsion to persistent circular motion, purely induced by memory of the medium~\cite{narinder2018memory}. Since the microswimmers move in a medium with very low Reynolds number, the influence of inertia on their dynamics is often ignored~\cite{narinder2018memory, samatas2023hydrodynamic, qiu2014swimming, wykes2016dynamic}. However, inertia is found to have unavoidable impact on the motion of
microswimmers~\cite{hubert2021scallop,deblais2018boundaries, fouxon2019inertial, pande2017setting, scholz2018inertial, manacorda2017lattice}.
Inertia can enhance (or hinder) the swimming speed of a microswimmer depending on whether the swimmer generates the thrust at the back or in front of its body~\cite{wang2012inertial}. Inertia can also influence the mean speed of microswimmers based on their rotation-translation coupling ~\cite{fouxon2019inertial}. 
 
In this letter, we theoretically demonstrate that even in the absence of any external torque or confinement, an inertial achiral microswimmer can make an unexpected transition from random self-propulsion to rotational (circular or elliptical) motion by tuning the inertial time or persistent duration of memory in the medium. Unlike the chiral microswimmers, here, the rotational motion undergoes spontaneous local direction reversals with a steady state angular diffusion with no specific direction of the rotational motion. Such an engrossing behaviour is seen by a minimal two-dimensional (2d) generalized Langevin model of non-Markovian dynamics of an inertial Ornstein-Uhlenbeck particle. Moreover, the swimmer with sufficient inertial memory gets confined or trapped with increase in duration of activity even in the absence of any external confinement.

In our model, we consider the 2d motion of an inertial microswimmer in a non-Newtonian fluid with a finite memory. 
The equation of motion of such a microswimmer is described by the generalized Langevin model of non-Markovian dynamics of an inertial active Ornstein-Uhlenbeck particle~\cite{sevilla2019generalized, sevilla2018nonequil} 
\begin{equation}
   m\ddot{\bf r}(t)=-\int_{0}^{t} g(t-t') \dot{\bf r}(t')\, dt' + \sqrt{D} \boldsymbol{\xi}(t).
     \label{eq:model-dynamics}
\end{equation}
Here, ${\bf r}(t) = x(t) \hat{i} + y(t) \hat{j}$ represents the position vector of the microswimmer, $m$ represents its mass, $\gamma$ is the viscous coefficient of the medium, and $D$ is the strength of the noise. 
The first term in the right hand side of Eq.~\eqref{eq:model-dynamics} is the viscous force with $g(t-t')$ as the memory kernel or viscous kernel, which is given by
\begin{equation}
    g(t - t') = \begin{cases}
\frac{\gamma}{t_c'}e^{-\frac{(t-t')}{t_c'}}; &  t\geq t'\\
0~~~~~~~~~; & t < t'.
\end{cases}
    \label{eq:mem-kern}
\end{equation}
The memory kernel $g(t-t')$ decays exponentially with time constant $t_c'$, where, $t_c'$ is the time that quantifies the persistent duration of memory in the medium or environment. The memory kernel often characterizes the viscoelastic nature of the non-Newtonian environment with a transient elasticity~\cite{sprenger2022active}. Here, the memory kernel follows the Maxwellian viscoelastic formalism as per which the elastic force of the fluid completely diminishes for sufficiently long interval of time~\cite{igor2012visco}. In the vanishing limit of $t_c'$, the memory kernel becomes a delta function [$g(t - t')=\delta(t-t')$], as a result, the suspension entirely turns into a viscous medium.
The term $\boldsymbol{\xi}(t)$ in Eq.~\eqref{eq:model-dynamics} is the induced active force or self-propulsion force, that follows the Ornstein-Uhlenbeck process
\begin{equation}
t_c\dot{\boldsymbol{\xi}}(t) = -\boldsymbol{\xi}(t) + \boldsymbol{\eta}(t),
\label{eq:noise}
\end{equation}
with the statistical properties 
\begin{equation}
    \langle \xi_i(t) \rangle = 0\quad \text{and} \quad \langle \xi_i(t) \xi_j(t') \rangle = \frac{\delta_{ij}}{2 t_c} \exp\left(\frac{|t - t'|}{t_c}\right).
\end{equation}
Here, $t_c$ represents the self-propulsion time or activity time, i.e., the time up to which the microswimmer has a tendency to self-propel in the medium. $\boldsymbol{\eta}(t)$ is a delta-correlated Gaussian white noise and $(i,j) \in (x,y)$. 
When $t_c = t_c'$, the generalized fluctuation dissipation relation is validated and the system approaches thermal equilibrium for $D=2\gamma k_{B} T$ with $T$ representing the temperature of the system and $k_{B}$ as the Boltzmann constant~\cite{prost2009generalized}.

The mean displacement (MD) of the microswimmer can be defined as
\begin{equation}
    \langle \Delta{\bf r}(t) \rangle = \langle {\bf r}(t) - {\bf r}(0) \rangle,
    \label{eq:md}
\end{equation}
with ${\bf r}(0)$ representing the initial position of the swimmer. The orientation of the swimmer can be determined by the direction of its instantaneous velocity. In order to describe the orientation of the swimmer, we introduce a parameter $\theta$ such that $\theta = \tan^{-1} \left( \frac{v_y}{v_x} \right)$ for all $\theta \in \left[ -\pi,\pi \right]$, with $v_x$ and $v_y$ being the $x$ and $y$ components of the instantaneous velocity ${\bf v}$ of the swimmer, respectively. Here, $\theta$ is the angle that the instantaneous velocity of the swimmer makes with the $x$ axis and the time evolution of $\theta$ quantifies the instantaneous angular velocity. Hence, the average angular velocity  $\langle \omega \rangle$ can be defined as $\left\langle \frac{d\theta}{dt} \right\rangle$ ~\cite{narinder2018memory}. 

\begin{figure}
    \centering
   \includegraphics[width=1\linewidth]{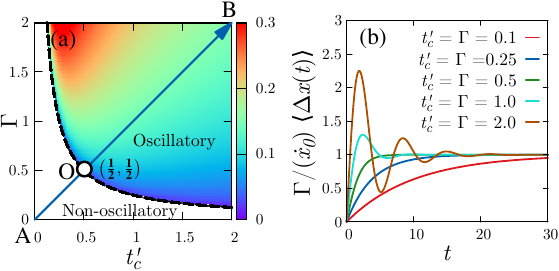}
    \caption{(a) Phase diagram separating the oscillatory and non-oscillatory regimes in the $t_c'-\Gamma$ parameter space [Eq.~\eqref{eq:soln_free}]. The color map in (a) shows the evolution of frequency of oscillation with $\Gamma$ and $t_c'$. (b) $\frac{\Gamma}{\dot{x}_0} \langle \Delta x(t) \rangle$ as a function of $t$ [Eq.~\eqref{eq:md_free_nomag}] for different values of $\Gamma$ and $t_c'$ along the diagonal line AB of the phase diagram in (a). The other common parameters are $m = t_c = \dot{x}_0 = D =  1$, and $x_0 = 0$.}
    \label{fig:oscillation_free}
\end{figure}
\begin{figure*}[!ht]
    \centering
   \includegraphics[width=0.75\linewidth]{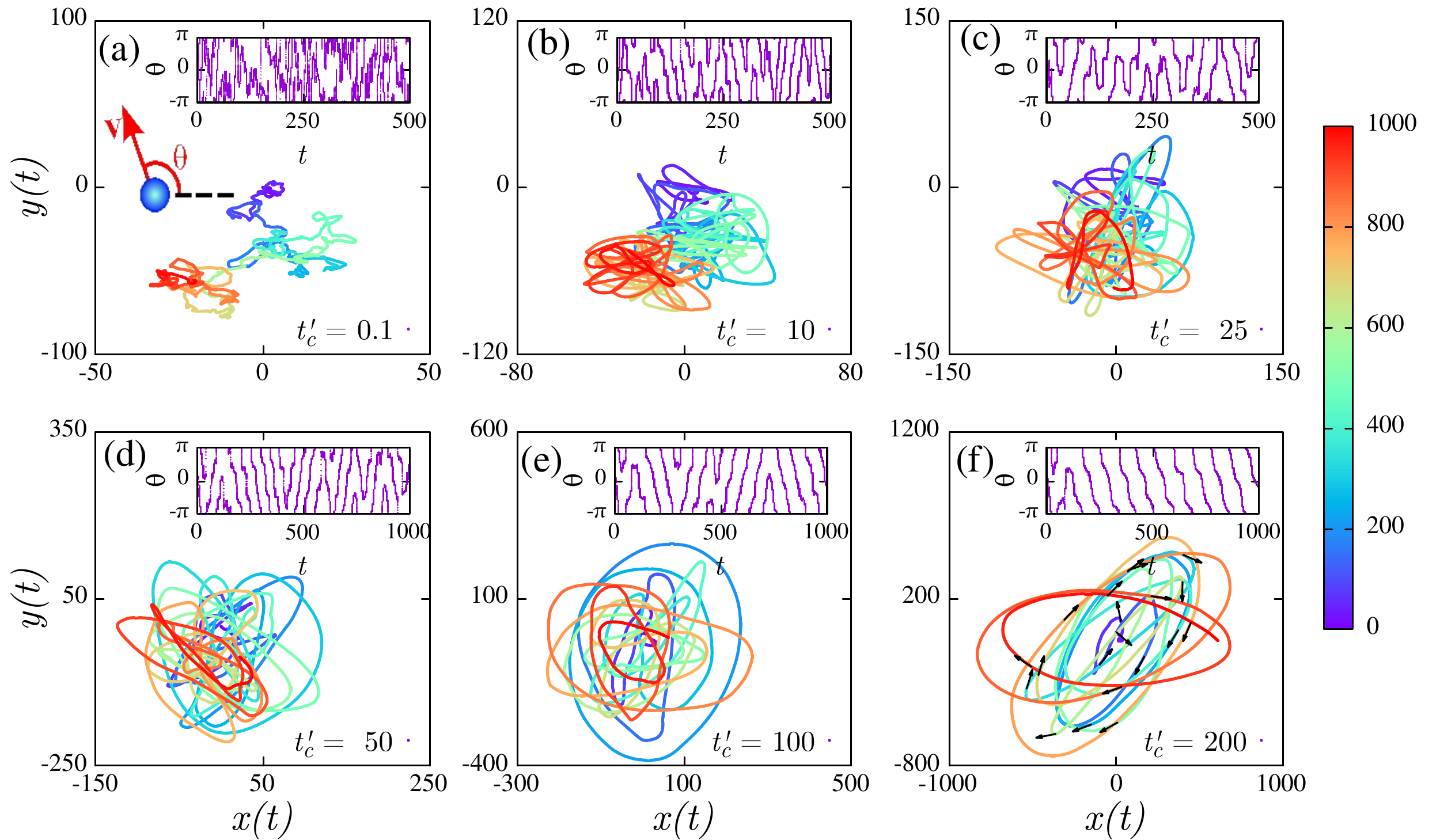}
    \caption{Simulated trajectories of a microswimmer for different values of $t_c'$ are plotted  in (a) for $t_c'=0.1$, (b) for $t_c'=10$, (c) for $t_c'=25$, (d) for $t_c'=50$, (e) for $t_c'=100$, and (f) for $t_c'=200$. The insets of (a)-(f) shows the evolution of its corresponding orientation $\theta$ with $t$. The other fixed parameters for all the plots are $m = \gamma = D = t_c = 1$, ${\bf r}_0=0\hat{i} + 0\hat{j}$, and ${\bf v}_0=1\hat{i} + 1\hat{j}$. The colour map shows the time evolution of the trajectories and the arrows in (f) represent the instantaneous orientation of the swimmer. A schematic diagram for the instantaneous orientation ($\theta$) of the swimmer is depicted in the inset of (a).}
    \label{fig:traj_free_tcp}
\end{figure*}
By performing Laplace transform on Eq.~\eqref{eq:model-dynamics}, the exact solution of Eq.~\eqref{eq:model-dynamics} with initial conditions ${\bf r}(0) = {\bf r}_0 [ = (x_0\hat{i} + y_0\hat{j})]$ and ${\bf \dot{r}}(0) = {\bf v}_0 [ = (\dot{x}_0\hat{i} + \dot{y}_0\hat{j})]$ can be obtained as
\begin{equation}
  {\bf r}(t)=  {\bf r}_0 +  \sum_{i=0}^{2} a_i \biggr[{\bf v_0} e^{s_i t}+ \frac{\sqrt{D}}{m} \int_{0}^{t} e^{s_i(t-t')}\boldsymbol{\xi}(t') \,dt'\biggr].
  \label{eq:soln_free}
\end{equation}
Here, $s_i$'s are given by the expression
\begin{equation}
s_0=0 \quad\text{and}\quad 
s_{1,2} = -\frac{1}{2} \left( \frac{1}{t_c^\prime} \pm \frac{\sqrt{1 -4t_c^\prime \Gamma}}{t_c^\prime} \right).
\label{eq:si}
\end{equation}
The coefficients $a_i$'s are given  by
\begin{equation}   
        a_0=\frac{1}{\Gamma}\quad \text{and} \quad a_{1,2} = \frac{\pm t_c'-t_c'\sqrt{1-4 \Gamma  t_c'}}{\pm\left(1-4 \Gamma  t_c'\right)+\sqrt{1-4 \Gamma  t_c'}},
\end{equation}
with $\Gamma = \frac{\gamma}{m}$ being the inverse inertial time scale of the swimmer. From the exact solution of the dynamics [Eq.~\eqref{eq:soln_free}], we establish a phase diagram in the inertia-memory ($t_c'$, $\Gamma$) parameter space [see Fig.~\ref{fig:oscillation_free}(a)] separating the oscillatory and non-oscillatory behaviour of motion of the swimmer. The oscillatory and non-oscillatory motions of the swimmer are separated by a phase boundary $t_c'\Gamma = \frac{1}{4}$. The microswimmer experiences oscillations in its motion for the phase with $t_c'\Gamma > \frac{1}{4}$ and no oscillation in its motion for the phase with  $t_c'\Gamma < \frac{1}{4}$ in the $t_c'-\Gamma$ parameter space. The exact calculation of frequency of oscillation shows its dependence on $\Gamma$ and $t_c'$. It decreases with increase in $t_c'$, increases with increase in $\Gamma$, and remains same along the diagonal line, i.e., for $\Gamma=t_c'$ [see the oscillatory regime of Fig.~1(a)]. In order to distinguish the oscillatory and non-oscillatory phases, in Fig.~\ref{fig:oscillation_free}(b), we have shown the time evolution of the $x$-component of mean displacement of the swimmer corresponding to all the points along the diagonal line AB of the $t_c'-\Gamma$ phase diagram [in Fig.~\ref{fig:oscillation_free}(a)], by plotting $\frac{\Gamma}{\dot{x}_0} \langle \Delta x(t) \rangle$ versus $t$. Using Eq.~\eqref{eq:md}, the MD at any instant of time is calculated as 
\begin{equation}
    \langle \Delta {\bf r}(t) \rangle =
    \frac{{\bf v}_0}{\Gamma} +{\bf v}_0 \left(a_1 e^{s_1t}+a_2 e^{s_2t} \right).
    \label{eq:md_free_nomag}
\end{equation}
The steady state MD is found to be $\langle \Delta{\bf r} \rangle_{st} = \lim\limits_{t\to\infty}\langle \Delta{\bf r}(t) \rangle  = \frac{{\bf v_0}}{\Gamma}$. While passing from the non-oscillatory to the oscillatory phase along the AB line in the $t_c'-\Gamma$ phase diagram, if the chosen parameters $(t_c', \Gamma)$ lie above the point $O (\frac{1}{2}, \frac{1}{2}$), the time evolution of corresponding MD is oscillatory and if these points lie below the point $O (\frac{1}{2}, \frac{1}{2}$), the time evolution of corresponding MD is non-oscillatory [see Fig.~\ref{fig:oscillation_free}(b)]. Moreover, the intermediate time regime of MD is oscillatory. Similar observation is noted from the mean-square displacement (MSD) calculation [See the Supplementary Material (SM)~\cite{supplementary}] .

\begin{figure}[!ht]
    \centering
    \includegraphics[width=\linewidth]{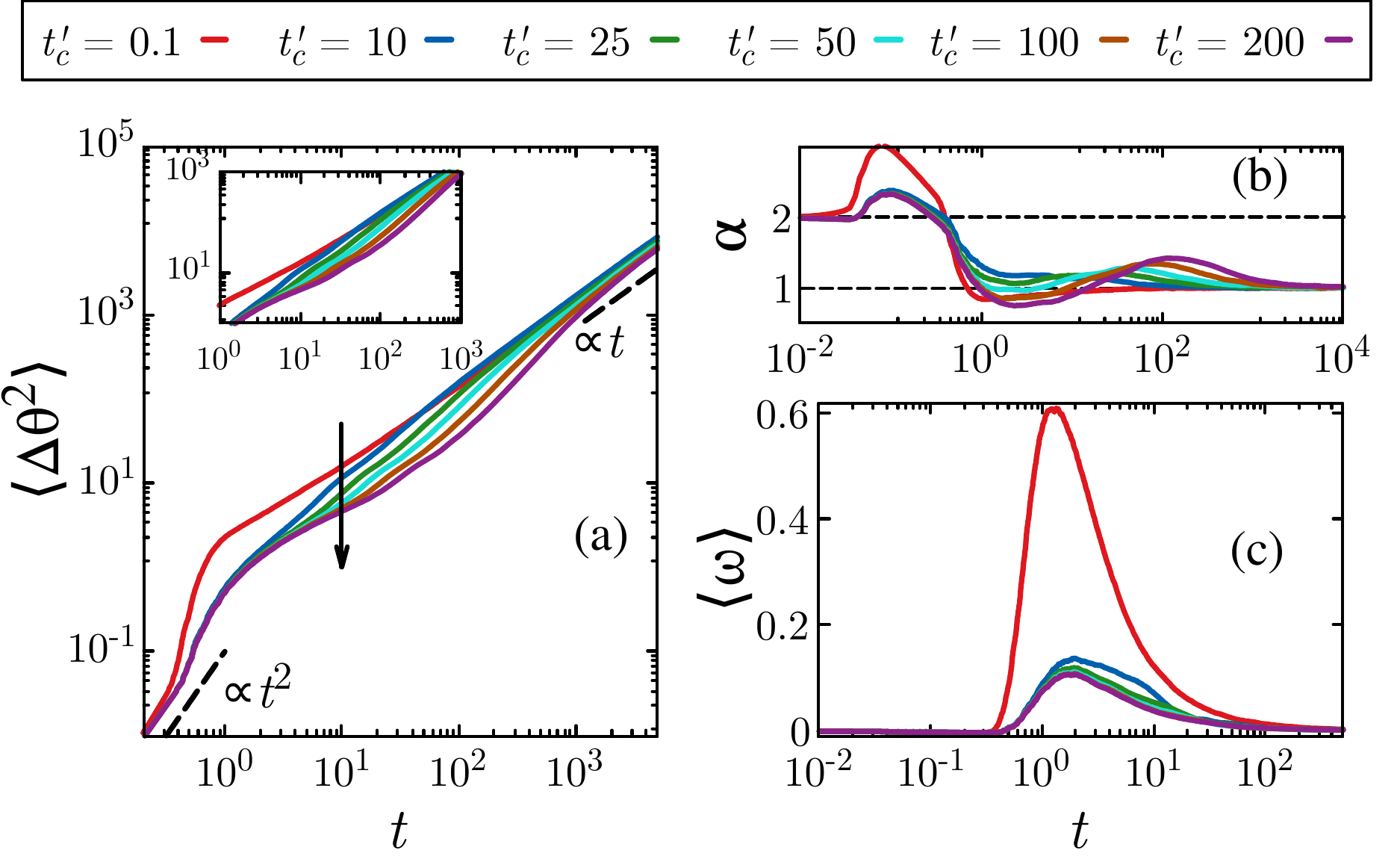}
    \caption{(a) Log-Log representation of the MSAD vs $t$ of the swimmer for different values of $t_c'$. The downward arrow points the increasing value of $t_c'$. The expanded view of the intermediate time regimes of MSAD is shown in inset of (a). (b) $\alpha$ vs $t$ for different values of $t_c'$. (c) $\langle \omega \rangle$ vs $t$ for different values of $t_c'$. The other common parameters are $D = t_c = \gamma = m = 1$, ${\bf v_0} = 1\hat{i} + 1\hat{j}$, and ${\bf r_0} = 0\hat{i} + 0\hat{j}$.}
    \label{fig:omega_t}
\end{figure}
The simulation of the dynamics [Eq.~\eqref{eq:model-dynamics}] is carried out using second order modified Euler method~\cite{kloeden1992numerical} and Fox algorithm~\cite{fox1988fast}. The simulation is run up to $10^3$ time steps and the numerical integration is performed with a time step of $10^{-3}$. Averages of the physical quantities are measured over $10^4$ realizations. 
By analyzing the instantaneous 2d trajectories of the microswimmer in the $xy$-plane [see Fig.~\ref{fig:traj_free_tcp}], it is interestingly observed that by fine tuning the persistent duration of memory or (and) inertial time in the oscillatory regime of $t_c'-\Gamma$ parameter space, the swimmer performs rotational (circular or elliptical) motion even in the absence of any external torque or confinement. The emergence of this behaviour is only due to the presence of inertial memory or elastic nature of the fluid. 

Figures~\ref{fig:traj_free_tcp}(a)-\ref{fig:traj_free_tcp}(f) present the instantaneous 2d trajectories of a microswimmer for different values of $t_c'$, keeping $\Gamma$ fixed. In the insets of Figs.~\ref{fig:traj_free_tcp}(a)-(f), we also present the data corresponding to the time evolution of orientation ($\theta$) of the swimmer. For $t_c'=0.1$, the instantaneous trajectory [Fig.~\ref{fig:traj_free_tcp}(a)] exhibits almost random behaviour, reflecting random self-propulsion of the swimmer and the time evolution of corresponding $\theta$ [inset of Fig.~\ref{fig:traj_free_tcp}(a)] on an average shows a random pattern. However, with increase in $t_c'$ value (i.e. at around $t_c'=10$), the swimmer is influenced by the elastic nature of the suspension and hence tries to make smooth and curvy type trajectory [Fig.~\ref{fig:traj_free_tcp}(b)]. With further increase in $t_c'$ value, at around $t_c' = 25$ [see Fig.~\ref{fig:traj_free_tcp}(c)], the swimmer starts likely to form circular (or elliptical) trajectory. For higher values of $t_c'$, the trajectories adopt almost elliptical shape, as shown in Figs.~\ref{fig:traj_free_tcp} (d), (e), and (f) for $t_c'=50$, $t_c'=100$, and $t_c'=200$, respectively. Hence, the increase of $t_c'$ leads the swimmer to make an unforeseen transition from random self-propulsion to a different dynamical regime characterized by persistent rotation. 
The plots of corresponding $\theta$ versus $t$ [insets of Figs.~\ref{fig:traj_free_tcp}(d), (e), and (f)] also show slanted stripes which is the consequences of rotational motion, either in clockwise ( $\pi$ to $-\pi$) or counter-clockwise ($-\pi$ to $\pi$) direction. This observation reflects that during time evolution, such rotational orbits spontaneously reverse their orientations. The frequency of direction reversals is found to be same as the frequency of oscillation, that decreases with increase in $t_c'$. Remarkably, the occurrence of these orientation reversals become very uncommon as $t_c'$ increases. At the same time, the elliptical trajectories become more stable and are clearly identified, as seen from Fig.~\ref{fig:traj_free_tcp}(f). The time evolution of corresponding $\theta$ on an average becomes linear over a long interval of time [see inset of Fig.~\ref{fig:traj_free_tcp}(f)]. Similarly, keeping $t_c'$ fixed and varying $\Gamma$ as well as varying both $t_c'$ and $\Gamma$ simultaneously in the $t_c'-\Gamma$ parameter space, we observe the same attributes and our assessments based on the trajectory plots is supported by the MSD calculation [see the Supplementary Material (SM)~\cite{supplementary}]. 

The rotational motion of the swimmer is characterized by computing the mean-square angular displacement (MSAD) $\langle \Delta\theta^2 \rangle$ [with $\Delta \theta = \theta - \langle \theta \rangle$] which is plotted as a function of $t$ for different values of $t_c'$ in Fig.~\ref{fig:omega_t}(a). For all $t_c'$ values, the initial and long time regimes are found to be ballistic and diffusive, respectively, whereas the intermediate time regime does not follow either of these behaviours. The deviation from ballistic and  diffusive behaviours is found to be significant for higher values of $t_c'$ [see inset of Fig.~\ref{fig:omega_t}(a)]. In order to understand the behaviour of MSAD in the intermediate time regime, we introduce a parameter $\alpha$ such that $\langle \Delta\theta^2\rangle \sim t^\alpha$ and plot the exponent $\alpha$ versus $t$ for different values of $t_c'$ in Fig.~\ref{fig:omega_t}(b). In the intermediate time regime, $\alpha$ varies from $2.6$ to $0.8$, that confirms the motion to be oscillatory and varying between super-ballistic ($\alpha >2$) and sub-diffusive ($\alpha <1$) behaviours. However, the time asymptotic regime of rotational motion is always diffusive ($\alpha=1$). This indicates that with increase in $t_c'$ value, although swimmer performs circular or elliptical trajectory with a specific orientation, due to the stochastic change of direction of rotation, on an average there is no preferential orientation of the swimmer at the steady state. That is why, the mean angular velocity $\langle \omega \rangle$ of the rotational orbits vanishes in the time asymptotic limit [see Fig.~\ref{fig:omega_t}(c)]. In Fig.~\ref{fig:omega_t}(c), we plot $\langle \omega \rangle$ as a function of $t$ for different values of $t_c'$. In the intermediate time regime, $\langle \omega \rangle$ is finite and found to be non-monotonic with $t$. For a particular value of $t_c'$, initially it increases, features a maximum, then decreases, and approaches zero in the time asymptotic limit. The magnitude of $\langle \omega \rangle$ at the maximum position decreases with increasing $t_c'$ and gets saturated above a certain value of $t_c'$ at which the swimmer starts performing rotational motion. The rotational motion occurs at the intermediate time regimes, where the motion is oscillatory and hence both inertia and memory are responsible for the rotational motion.

\begin{figure}
    \centering
    \includegraphics[width=0.9\linewidth]{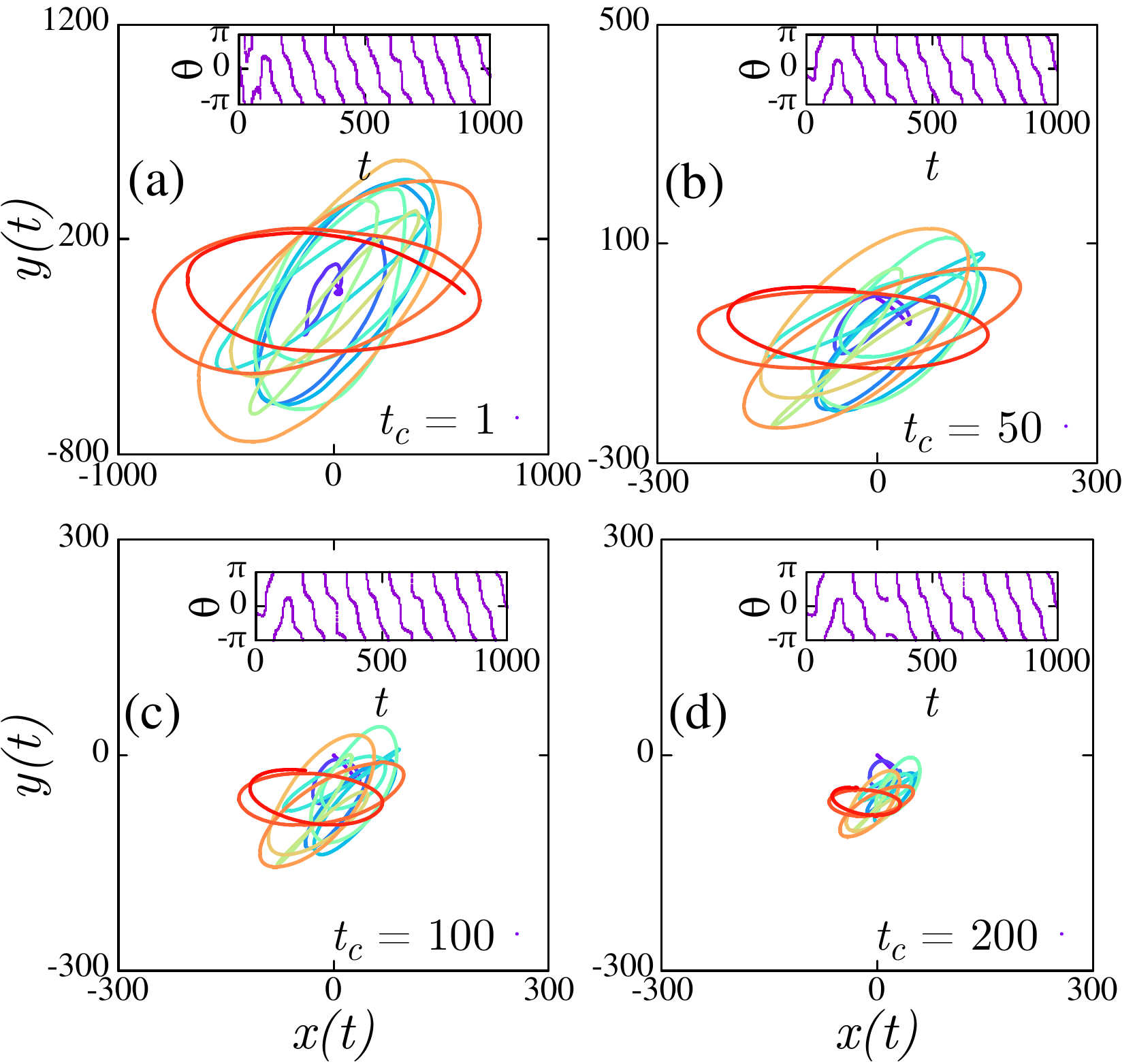}
    \caption{Instantaneous swimmer trajectories for different values of $t_c$ are plotted in (a)  $t_c=1$, (b) $t_c=50$, (c) $t_c=100$, and (d) $t_c=200$, for fixed $t_c'=200$. Insets: Corresponding time evolution of their instantaneous orientation. The other common parameters are $m = \gamma = D = 1$, ${\bf v_0} = 1\hat{i} + 1\hat{j}$, and ${\bf r_0} = 0\hat{i} + 0\hat{j}$.}
    \label{fig:traj_tc}
\end{figure}

Besides, fixing $t_c'=200$ at which the elliptical trajectories are stabilized, if we further increase the value of $t_{c}$, the elliptical orbits get compressed (see Fig.~\ref{fig:traj_tc}). This is a clear indication of trapping or confinement of the swimmer (with inertial memory) in the medium even in the absence of any external confinement. The presence of memory or elasticity in the medium provides a kind of harmonic confinement for the swimmer, as a consequence the swimmer with inertial memory performs rotational trajectory like an inertial active walker in a harmonic confinement \cite{dauchot2019dynamics}. Further, the compression of trajectory or trapping of the swimmer with increase in duration of activity complements the above feature as in Ref.~[\onlinecite {takatori2016acoustic}]. 

In conclusion, we have theoretically probed the self-propulsion of an inertial-activated microswimmer in a non-Newtonian environment with a finite memory. We investigate the motion of the swimmer by solving the generalized Langevin dynamics  of an inertial active Ornstein-Uhlenbeck particle. The obtained results surprisingly uncover a transition from random self-propulsion to rotational (circular or elliptical) motion upon fine tuning of the inertial time scale or persistent duration of memory in the medium. This rotational motion is followed by spontaneous local orientation reversals. Further, the average angular velocity of the swimmer has a finite value in the transient limit and zero value in the time asymptotic limit or at steady state. This implies that at steady state, the angular motion is diffusive and on an average there is no preferential orientation of the rotational motion. Moreover, these rotational orbits of the swimmer appear only in the oscillatory regime of the inertia-memory parameter space. Unlike the synthetic microswimmer with shape asymmetry~\cite{kummel2013circular}, here, the emergence of circular or elliptical orbits is due to the presence of inertial memory in the medium. Interestingly, the swimmer with sufficient inertial memory gets trapped with increase in duration of activity even in the absence of any external confinement. In the absence of either inertia or memory, the rotational motion does not occur [see SM~\cite{supplementary}].

We acknowledge the 8th statphysics community meeting (ICTS/ISPCM2023/02), during which we had lots of fruitful discussions. MS acknowledges the start-up grant from UGC, SERB-SURE grant (SUR/2022/000377) from DST, Govt. of India for financial support and thanks to Stefan Klumpp for critical reading of the manuscript and many useful comments and Debasish Choudhury for useful discussions and suggestions. We also thank Tapan Mishra and Ramesh Nath for reading of the manuscript and their comments.

\textsuperscript{$\dag$} Authors contributed equally to this work
%
\end{document}